\documentstyle[12pt]{article}
\begin{document}
\input{feynman}
\begin{titlepage}
\begin{centering}
{\large{\bf Static Quantities of the W Boson
in the MSSM}}\\
\vspace{.5in}
{{\bf A. B. Lahanas } \hspace{.1cm}}$^{\dag}$ \\
\vspace{.2in}
University of Athens, Physics Department,
Nuclear and Particle Physics Section\\
\vspace{.05in}
Ilissia, GR - 15771  Athens, GREECE\\
\vspace{.2in}
and \\
\vspace{.2in}
{{\bf V. C. Spanos } \hspace{.1cm}}$^{\ddag}$  \\
\vspace{.2in}
NRCPS  `Democritos',
Institute of Nuclear Physics\\
Aghia Paraskevi, GR - 15310\,Athens, GREECE\\
\vspace{.4in}
{\bf Abstract}\\
\vspace{.1in}
\end{centering}
{
\noindent
We systematically analyze the anomalous dipole $\Delta k_{\gamma}$
and quadrupole $\Delta Q_{\gamma}$ moments of the
W gauge bosons in the context of the minimal supersymmetric standard model
as functions of the soft SUSY breaking parameters
$A_0,m_0,M_{1/2}$  and the
top quark mass. The severe constraints imposed by the radiative
breaking mechanism of the electroweak symmetry\,$ SU(2)\times U(1)$\, are duly
taken into account. The supesymmetric values of $\Delta k_{\gamma}$
and $\Delta Q_{\gamma}$ can be largely different, in some cases, from the
standard model predictions but of the same order of magnitude for values  of
\quad $A_0,m_0,M_{1/2} \leq {\cal O}(1 TeV) $. Therefore possible
supersymmetric structure can be probed provided the accuracy of measurements
for $\Delta k_{\gamma}$, $\Delta Q_{\gamma}$ reaches $10^{-2}-10^{-3}$
and hence hard to be detected at LEP2. If deviations from the standard model
predictions are observed at LEP2, most likely these are
not due to an underlying supersymmetric structure. In cases where
$M_{1/2} \ll  A_0,m_0 $, the charginos and neutralinos may give substantial
contributions saturating the LEP2 sensitivity limits. This occurs when their
masses $m_{\tilde C},m_{\tilde Z}$ turn out to be both light satisfying
$m_{\tilde C}+m_{\tilde Z} \simeq M_W $. However these extreme cases
are perturbatively untrustworthy and besides unnatural
for they occupy a small region in the parameter space. }

%
\par
\vspace{4mm}
\begin{flushleft}
Athens University\\
UA/NPPS - 15/1994\\
April 1994\quad(revised)\\
hep-ph/9405298 \\
\end{flushleft}
\rule[0.in]{4.5in}{.01in}\\
E-mail:
$^{{\hspace{.2cm}}\dag}$ { \hspace{.1cm}
alahanas@atlas.uoa.ariadne-t.gr} ,\quad
$^{{\hspace{.2cm}}\ddag}$ { \hspace{.1cm}
spanos@cyclades.nrcps.ariadne-t.gr}
\end{titlepage}
%
One of the most crucial tests of the standard theory of electroweak
interactions will be the study of the three gauge boson couplings to
be probed in forthcoming experiments in the near or remote future.
Although there is little doubt that the nonabelian structure of the
standard model is the right framework for describing the electroweak phenomena
at low energies, in the vicinity of the electroweak scale, nevertheless
we are still lacking a direct experimental verification of it. Such a study
will take place at LEP2 as well as in next round experiments at HERA, NLC,
LHC, with high accuracy putting bounds on the dipole  $\Delta k_{\gamma}$  and
quadrupole  $\Delta Q_{\gamma}$  form factors which are directly related to
the anomalous
magnetic moment  $\mu_{W}$  and the electric quadrupole moment $Q_{W}$
of the W\,-boson $^{\cite{gounar}}$.
The standard model predicts   $\Delta k_{\gamma}=\Delta Q_{\gamma}=0$ at the
tree level but higher order corrections modify these values by finite
amounts that can be tested in the laboratory with high accuracy of the
order of $10^{{-2}}$ to $10^{{-3}}$  $^{\cite{baur}}$. Such measurements can
be of vital importance not only for the self consistency of the standard
model but also for probing possible structure beyond that of the standard
theory signalling the presence of new physics.
\par
The last years there has been a revived interest towards supersymmetric
extensions of the standard model. The supersymmetric structure seems to be
a necessary ingredient in the efforts towards embedding the standard model
in larger schemes unifying all existing forces of nature and is also suggested
by precision data on the gauge couplings $\alpha_{{1}}, \alpha_{{2}}
, \alpha_{{3}}$  which merge at a unification scale
$M_{GUT} \simeq 10^{ {16}}$ GeV,
provided the SUSY breaking \mbox{scale} $M_{S}$  lies in the TeV
range $^{\cite{ekn,abf}}$. Such relatively low values for the supersymmetry
breaking scale $ \simeq $ TeV may have important consequences for
phenomenology.
The TeV scale may be the onset of new physics and detection of supersymmetric
particles with masses $\leq O(M_{S})$  might not be out of reach in
 future experiments. For this we need energies and luminocities
that production of new particles is feasible with rates that are accessible to
the new machines. Therefore phenomenological study of supersymmetry is of
utmost importance. Below the threshold for the production of superpartners of
the known particles the only evidence for the existence of supersymmetry
will be the study of physical quantities which are affected by the presence
of the underlying supersymmetric structure. In this case
the supersymmetric particles are not observed
in the final states, and hence we talk about {\em virtual SUSY}; however
they induce radiative corrections to the physical quantities
of interest
and make them deviate from their standard model values. In order to measure
these one needs, at the theoretical level, higher order calculations while
experimentally we need high accuracy tests capable of measuring such small
differences.\,In case the experiments point towards the affirmitive it
will be an evidence for the existence of new physics SUSY being a
strong candidate to play that role.

In this paper we study the radiative corrections to the dipole and quadrupole
moments of the W - bosons in the minimal supersymmetric standard model (MSSM)
with soft supesymmetry breaking terms.\,It is a well known fact that softly
broken supersymmetry leads to $SU(2)_L \times U(1)_Y$ symmetry breaking
through radiative effects $^{\cite{ir,iban,ikkt,klnq}}$ and therefore within
the MSSM the elegant ideas of supersymmetry, gauge coupling unification and
natural explanation of the hierarchy
${M_W}/{M_{Planck}} \simeq 10^{{-16}}$ can be sumiltaneously realized.

There are numerous papers studying the phenomenology of
$\Delta k_{\gamma},\Delta Q_{\gamma}$
both on and off shell in the context of the
standard model (SM)  $^{\cite {bard}}$. Also supersymmetric versions of the SM
have been considered in which SUSY is either exact $^{\cite {proy}}$ or
broken $^{\cite {ali,hewett}}$ by soft terms but there is no systematic, to
our knowledge, phenomenological analysis that properly takes into account all
the effects of supersymmetry breaking and the constraints imposed by the
the renormalization group and the
radiative breaking of the electroweak symmetry
\footnote { Even in ref.${\cite {ali}}$
the SUSY breaking effects are actually ignored since only the supersymmetric
limit of the MSSM is considered in the discussion
of the physical results.}.
In this work we undertake this problem
and discuss the supersymmetric values of the dipole and quadrupole moments
as functions of the soft SUSY breaking terms and the top quark mass,
taking into account all the constraints imposed by the radiative breaking
of the electroweak symmetry. As a sneak preview of our results we state that
the supersymmetric values differ substantially from those of the standard
model, in some cases, contrary to what hes been claimed in
the literature $^{\cite {ali}}$.

The MSSM is described by a lagrangian
\begin{equation}
{\cal L}={\cal L}_{{SUSY}}+{\cal L}_{{soft}}
\end{equation}
where $ {\cal L}_{{SUSY}}$ is its supersymmetric part derived from
a superpotential ${\cal W}$ bearing the form  $^{\cite {nilles}}$
\begin{equation}
{\cal W}=(h_U \hat{Q}^i \hat{H}_2^j \hat{U}^c
         + h_D \hat{Q}^i \hat{H}_1^j \hat{D}^c
         + h_E \hat{L}^i \hat{H}_1^j \hat{E}^c
         + \mu \hat{H}_1^i \hat{H}_2^j) \epsilon_{ij} \quad,\quad
          \epsilon_{12}=+1
\end{equation}
and  $ {\cal L}_{{soft}}$ is its supersymmetry breaking part given by
\begin{eqnarray}
- {\cal L}_{{soft}}&=&{\sum_{i}} m_i^2 |\Phi_i|^2
         + (h_U A_U Q H_2 U^c
         + h_D A_D Q H_1 D^c
         + h_E A_L L H_1 E^c+h.c.)\nonumber\\
                         &+&(  \mu B H_1 H_2 +h.c.)
         +\frac {1}{2} \sum_a M_a {\bar{\lambda}}_a \lambda_a .
\end{eqnarray}
In Eq. (3) the sum extends over all scalar fields involved and we have
suppressed  all familly indices.
\par
In our analysis we assume universal boundary conditions for the soft
masses at a unification scale
$M_{ {GUT}} \simeq 10^{ 16}$ GeV. The evolution of all couplings
as well as all soft masses and the mixing parameters $\mu , B$ from
$M_{ {GUT}}$ down to energies $E$ in the vicinity of the electroweak
scale is given by their renormalization group equations (RGE)
$^{\cite {ikkt,nilles}}$~. These are known up to two loop
order    $^{\cite {mv}}$.
\,As arbitrary parameters of the model we take the running top quark mass
$m_t(M_{ Z})$\footnote{ The physical top quark mass $M_t$ is defined
by $M_t=m_t(M_t)/(1+{{ 4\alpha_3(M_t)}/{3\pi}}) $
when the one loop corrections are taken into account
$^{\cite {gray}}$       . }
at the Z\,-boson mass the angle  $\beta$ defined by
$\tan{\beta}={v_{ 2}(M_{ Z})/v_{ 1}(M_{ Z})}$
and the soft SUSY breaking parameters
$A_{ 0},m_{ 0}, M_{ {1/2}}$ at \mbox{$M_{ {GUT}}$.}
$v_{ {1,2}}(M_{ Z})$ are the vacuum expectation values of the Higgses
$H_{ {1,2}}$ and the values of the trilinear scalar couplings
$A_0$, common scalar mass $m_0$ and common
gaugino mass  $M_{1/2}$ are meant at the unification scale
$M_{ {GUT}}$. In this approach, which has been adopted by other authors
too $^{\cite {ramon,arno}}$   $B,\mu$ are not free parameters; in fact their
values at   $M_{ Z}$ are determined through the minimizing equations
of the scalar potential,
\newpage
\begin{eqnarray}
\frac {M_{ Z}^{ 2}}{2}&=&\frac {\bar {m}^{ 2}_{ 1}-
   \bar {m}^{ 2}_{ 2}  tan^{ 2}\beta}
   {tan^{ 2}\beta- 1}  \label{min1}    \\
                         sin2\beta&=&-\frac {2B\mu}
                        {\bar {m}^{ 2}_{ 1}
                        +\bar {m}^{ 2}_{ 2}}\quad\quad . \label{min2}
\end{eqnarray}
In Eqs. (\ref{min1}) and (\ref{min2})  the masses appearing are defined by
\begin{equation}
\bar {m}^{ 2}_{ {1,2}}={m}^{ 2}_{ {1,2}}+
\frac {\partial \Delta V}{\partial v_{ {1,2}}^{ 2}} \quad,
\quad  {m}^{ 2}_{ {1,2}}={m}^{ 2}_{H_{ {1,2}}}+
\mu^{ 2}
\end{equation}
where $\Delta V$ accounts for the one loop corrections to the
effective potential which should be included in the minimizing equations.
If they are not the results are known to depend drastically on the choice of
the scale becoming ambiguous and untrustworthy $^{\cite {zwi,arno,ramon}}$.
Eqs. (\ref{min1}) and (\ref{min2}) cannot be analytically solved to
obtain the values $B(M_{ Z}),\mu(M_{ Z})$ and our numerical routines
have to run several times to reach convergence. In solving the RGE's we
have also to properly take into account the appearance of the thresholds
of the various particles opened as we approach low energies. We have duly
taken care of these effects along the lines of ref $\cite {ross}$. We have
found that their presence little upsets the picture as long as one loop
corrections to the  $\Delta k_{\gamma}$ and  $\Delta Q_{\gamma}$ are
concerned.

The supersymmetric limit of the model is realized when all soft SUSY breaking
terms vanish, $A_0=m_0=M_{1/2}=0$,  the v.e.v's $v_{ 1}$
and $v_{ 2}$   become equal and the mixing parameter $\mu$  vanishes.
It is in this limit that particles and their superpartners get a common mass.
In that limit the quadrupole moment  $(\Delta Q_{\gamma})_{ SUSY}$
vanishes but the same does not happen for the dipole moment
$(\Delta k_{\gamma})_{ SUSY}$. For the latter the difference
$\Delta k_{\gamma}-(\Delta k_{\gamma})_{ SUSY}$ is a  measure of the
importance of
SUSY breaking effects on the dipole moment.

After this introductory remarks concerning the MSSM and an outline of the
numerical proceedure we shall follow we embark on discussing separately the
contributions of the various particles involved to the quantities of
interest.

The most general $ {W^+}{ W^-}{\gamma}$ vertex consistent with current
conservation can be written as
\begin{eqnarray}
\Gamma_{\mu \alpha \beta}& = &-ie \, \{ \,  f[2g_{\alpha \beta}{\Delta_\mu}+
       4(g_{\alpha \mu}{Q_\beta}-g_{\beta \mu}{Q_\alpha})]+ \nonumber\\
   &   &2 \, {\Delta k}_\gamma \,
                          (g_{\alpha \mu}{Q_\beta}-g_{\beta \mu}{Q_\alpha})+
       4 \, {\frac  {{\Delta Q}_\gamma}  {M_{ W}^{ 2}}} \, {\Delta_\mu}
       ({Q_\alpha}{Q_\beta}-{\frac {Q^{ 2}}{2}} g_{\alpha \beta})\}+...
\end{eqnarray}
where the  W's are on their mass shell and the ellipsis denote C, P violating
terms. The labelling of the momenta and the assignment of Lorentz indices
is as shown in \mbox{Figure 1.}
To lowest order $f=1, {\Delta k}_\gamma
={\Delta Q}_\gamma=0$. The values of the form factors
$ {\Delta k}_\gamma, {\Delta Q}_\gamma$
at zero momentum transfer are related to the actual magnetic dipole moment
$\mu_{ W}$   and electric quadrupole moment $Q_{ W}$
by $^{\cite {gounar,baur,bard}}$
\begin{eqnarray}
\mu_{ W}={\frac {e}{2M_{ W}}}\,(1+\kappa_{\gamma}+\lambda_{\gamma})
& , &Q_{ W}=-\,{\frac {e}{M_{ W}^2}}\,
(\kappa_{\gamma}-\lambda_{\gamma})
\nonumber
\end{eqnarray}
where $ {\Delta k}_\gamma \equiv \kappa_{\gamma}+\lambda_{\gamma}-1$ and
$ {\Delta Q}_\gamma \equiv -2 \, \lambda_{\gamma} $.

The standard model one loop predictions for  the dipole and quadrupole
moments have been calculated and can be traced in the literature
$^{\cite {bard}}$ .\, The contributions of gauge bosons and matter fermions of
the MSSM are identical to those of the SM. These are displayed in Table I in
units of ${g^2}/{16 {\pi ^2}}$.
For the contributions of the fermions we agree with the
findings of other authors as far as the isospin $T_{ 3}=-1/2$ fermions
are concerned but we disagree in the sign of the $T_{ 3}=+1/2$
chiral fermion contributions. This means for instance that ``up" and
``down" quarks' contributions to the dipole/quadrupole moments have the same
sign despite the fact that they carry opposite electric charges. This is due
to the fact that the triangle graph with the up quark coupled to the photon
and the corresponding one with the down quark playing that role are crossed;
the dipole and quadrupole terms however change sign under crossing,
 resulting to a same sign contribution,
unlike the anomaly  which preserves its sign yielding
$\simeq Trace(Q)$. Since this is a rather delicate point which, we think, has
been overlooked in previous works we shall discuss it in more detail.

The fermionic Lagrangian relevant for the calculation of  $\Delta k_{\gamma}$
and  $\Delta Q_{\gamma}$ is
\begin{equation}
{\cal L} \,=\, {\frac {g}{\sqrt 2}}(\,W^+_{ \mu}
{\bar{\Psi}}_{ L}{\gamma^{\mu}}{T_+}{\Psi}_{ L}+h.c.)
\,+\,e\,A_{\mu}\,( {\bar{\Psi}}_{ L}Q {\gamma^{\mu}}{\Psi}_{ L})
\end{equation}
In this $ {\Psi}_{ L}$  is a column involving all left handed fermions,
${T_{\pm}}={T_{1}}{\pm} i{T_{2}}$ are weak isospin raising/lowering matrices
and $Q$ is the diagonal electric charge matrix.
The graphs we have to calculate are shown in Figure 2, where $p\,,\,p ^\prime $
are the momenta carried in by the external $W$'s. The Feynman integral
of the graph shown in Figure 2b has a momentum dependence which follows from
 that of Figure 2a under the interchange $\alpha \rightleftharpoons \beta$
and
$p \rightleftharpoons p^\prime $ or  eqivalently  $ {Q_{\mu}}\rightleftharpoons
{Q_{\mu}}, {\Delta_{\mu}} \rightleftharpoons {-\Delta_{\mu}} $.
One is easily convinced of that by explicitly writing down the expressions
of the two graphs shown in Figure 2. Therefore one has
\begin{eqnarray}
Graph (2a)\,&=&\,Tr(T_+ T_- Q)\,{V_{\mu\alpha\beta}}(Q,\Delta)\nonumber\\
Graph (2b)\,&=&\,Tr(T_- T_+ Q)\,{V_{\mu\beta\alpha}}(Q,-\Delta)\nonumber
\end{eqnarray}
where the tensor structure of ${V_{\mu\alpha\beta}}(Q,\Delta)$ including the
axial anomaly term is
\begin{equation}
{V_{\mu\alpha\beta}}(Q,\Delta)\,=\,{\alpha_{ 0}}
{\epsilon_{\alpha\beta\mu\lambda}}{\Delta^{\lambda}}+
{\beta_1}g_{\alpha \beta}{\Delta_\mu}+
{\beta_2} (g_{\alpha \mu}{Q_\beta}-g_{\beta \mu}{Q_\alpha})+
{\beta_3} {\Delta_\mu} {Q_\alpha}{Q_\beta}+...
\end{equation}
The first term is the anomaly and $\beta_{ {1,2,3}}$ are form factors
where the contributions to the charge renormalization, dipole and quadrupole
moments are read from. The anomaly term does not flip its sign under
$\alpha \rightleftharpoons \beta$,
$ {Q_{\mu}}\rightleftharpoons{Q_{\mu}},
{\Delta_{\mu}} \rightleftharpoons {-\Delta_{\mu}} $
but the remaining terms do. As a result we get for the sum of the two graphs
\begin{equation}
\,Trace(\,Q\,\{T_-,T_+\})
{\epsilon_{\alpha\beta\mu\lambda}}{\Delta^{\lambda}}+
\,Trace(\,Q\,[T_+,T_-])({\beta_1}g_{\alpha \beta}{\Delta_\mu}+...)  \label{tr}
\end{equation}
where for the sake of the argument we have suppressed the fermion mass
dependence entering into $\beta_{ {1,2,3}}$ by assuming that all fermions
have the same mass. In the standard model $\{T_-,T_+\}=constant\times{\bf 1}$
and the anomaly is proportional to $Trace(Q)$ which vanishes, a well known
result. The second term of Eq.(\ref{tr}) however yields a contribution
proportional
to $Trace(Q\,T_{ 3})$, when all fermions have the same mass. From this it
becomes obvious that ``up" and ``down" quarks's contributions to the dipole\
quadrupole moments have the same sign since they carry opposite isospin and
electric charges. Since the fermion masses are different the graphs of
Figure 2  yield, ignoring the anomaly term,
\begin{equation}
{\sum_{f}}({Q_f}{T_f^{ 3}}){V^{\alpha\beta\mu}}(Q,\Delta,{m_f^{ 2}}
,{m_{{f^\prime}}^{ 2}})
\end{equation}
where $f , f^\prime$ are left handed fermions  belonging to doublets of the
weak $SU(2)$. The explicit expressions for
$\Delta k_{\gamma}$, $\Delta Q_{\gamma}$ of each fermion doublet
${f \choose {f^\prime}}_{ L}$  are given in the Table I, where they are
expressed in terms of the dimensionless ratios ${r_{{ f,}{ {f^\prime}}}}
=(m_{ {f,{f^\prime}}}/M_{ W})^{ 2}$. The factor $C_g$ appearing in these
formulae is the color factor, one for the leptons and three for the
quarks.

The discussion of the sleptons and squarks is complicated by the fact that
the superpartners ${\tilde f}_{ L},{\tilde f}_{ L}^c$ of the
known fermions  $f_{ L},f_{ L}^c$ are not mass eigenstates .
This is the case for the third familly fermions and especially for the stops
${\tilde t}_{ L},{\tilde t}_{ L}^c$ due to the heaviness of the
top quark which results to large
${\tilde t}_{ L},{\tilde t}_{ L}^c$ mixings. The first two families
have small mass and  such mixings are not large. In the stop
sector the relevant mass matrix squared has the following form,
\begin{equation}
{\bf {\cal M}}^{ 2}_{\tilde t}\,=\,\left(\begin{array}{lllll}
m^{ 2}_{ {Q_{ 3}}}+m^{ 2}_{ t}
&   &    &    & m_{ t}(A+\mu cot\beta)
\\
\quad\quad+{M^{ 2}_{ Z}}(cos2\beta)({\frac {1}{2}}
                       -{\frac {2}{3}}sin^{ 2}{\theta_{ W}}) &  &  &   &\\
 &  &  &  &  \\
 m_{ t}(A+\mu cot\beta) &   &  &   &  m^{ 2}_{ {U^{ c}_{ 3}}}+m^{ 2}_{ t}\\
&   &    &   &\quad\quad+{M^{ 2}_{ Z}}(cos2\beta)
                      ({\frac {2}{3}}sin^{ 2}{\theta_{ W}})
\end{array}\right)
\end{equation}
%
which is diagonalized by a matrix ${\bf K}^{\tilde t}$, i.e.
${\bf K}^{\tilde t} {\bf {\cal M}}^{ 2}_{ {\tilde t}}
{{\bf K}^{\tilde t}}^{\bf T} =diagonal$.
Since the top mass is large ${\bf K}^{\tilde t}$ deviates substantially from
unity. In the same way the sbottom and stau mass matrices get diagonalized by
${\bf K}^{\tilde b},{\bf K}^{\tilde \tau}$. However in that case the
corresponding
${\tilde f}_{ L},{\tilde f}_{ L}^c$ mixings are not substantial.
The sfermion contributions to $\Delta k_{\gamma}$, $\Delta Q_{\gamma}$
are presented in the Table I\,. They are expressed in terms of the
diagonalizing matrices discussed previously  and the dimensionless ratios
$R_{{ \tilde f}_{ i}} =
(m_{{ \tilde f}_{ i}}/{M_{ W}})^{ 2}$,
where $i=1,2$ runs over the mass eigenstates. In the supersymmetric limit
the matrices  ${\bf K}^{{\tilde f}, {\tilde f}^\prime}$ corresponding to the
sfermion doublet   ${{\tilde f} \choose {{\tilde f}^\prime}}_{ L}$
are unit matrices and $m_{{ \tilde f}_{ {1,2}}}$
($m_{{{ \tilde f}^\prime}_{ {1,2}}}$) become equal to the fermion mass
$m_{ f} (m_{{ f}^\prime}) $.
In our numerical analysis we have taken only the top quark mass to be
nonvanishing and therefore the aforementioned discussion regards only the
stops. The inclusion of the bottom and tau masses little affects the
quantities     $\Delta k_{\gamma}$ , $\Delta Q_{\gamma}$  owing to the fact
that the induced mixings in the corresponding sbottom and stau sectors
are small. In the Table I we also give the dipole moment in the
supersymmetric limit discussed earlier. Notice that in that limit the
fermion/sfermion contributions  to the quadrupole moment of each familly
separately cancel against each other.

We turn next to discuss the Higgs sector. In the MSSM we have two doublets
of Higgs scalars $H_{ 1},H_{ 2}$ giving masses to ``down"
($ T_{ 3}=-1/2 $) and ``up" ($T_{ 3}=1/2$ ) fermions when they
develop nonvanishing v.e.v's as a result of electroweak
symmetry breaking. The relative strength of their v.e.v's  is parametrized
by the angle $\beta$ defined as
\mbox{$\tan{\beta}={v_{ 2}}/v_{ 1}$ }\,. These are
running parameters depending on the scale, as is the running Z\,-boson
mass ${m_{ Z}^{ 2}}={(g^{ 2}+{g^\prime}^{ 2})}
(  {v_{ 1}^{ 2}}  +   {v_{ 2}^{ 2}} )/2 $.
The physical Z\,-boson mass  is defined as the pole of the Z\,-boson
propagator occurring therefore at the point $M_{ Z} $ for which
${m_{ Z}}(M_{ Z})= M_{ Z}$. Experimentally we know that
$ M_{ Z}=91.2 GeV $\,.

As a result of the electroweak symmetry breaking we have five physical Higgses
$A,h_{ 0},H_{ 0},H_{\pm}$ with the following masses,
$$
\begin{array}{ll}
A\,(neutral)&:\, {m^2_A}= {m^2_1}+{m^2_2}   \\ \\
{H_0,h_0    }\,(neutral) &:\,
{m_{ H,h }^2}={\frac {1}{2}} \{  {(m^2_A+M_Z^2)}^2\pm
  \sqrt {   {(m^2_A+M_Z^2)}^2 -4{M_Z^2} {m^2_A} {cos^2}(2\beta) }
                                                                 \}  \\
  &  \\
{H_\pm}\,(charged) &: \, m^2_{H_\pm} =m^2_A+ M_W^2
\end{array}
$$
The lightest of the neutral Higgses  $ h_0$ has always a tree level mass
smaller than  $M_Z$. However it is known that radiative corrections due
to the top/stops can be large, due to the heaviness of the top quark
$^{\cite {higgs}}$ and shift its upper limit to values that can exceed
$M_Z$. In that case   $ h_0$  can escape detection at LEP2.
In the supersymmetric limit  $h_0$ and the pseudoscalar $A$ become massless,
$H_\pm$ has a mass $M_W$ and $H_0$ a mass $M_Z$.

The Higgs contributions to the dipole and quadrupole moments are shown
in the \mbox{Table I} as functions of the dimensionless parameters
${R_\alpha}={({m_\alpha}/M_W)}^2, \alpha=
{h_0,H_0,A,H_\pm}$  and a mixing angle $\theta$ ;
for  $\sin^2\theta =1$, \, $h_0$ becomes the standard model Higgs scalar
\footnote{ $sin^2 \,\theta\,=\,(m^2_A +{M^2_Z} {\sin^2}(2\beta)-m^2_h)/
(m^2_H\,-\,m^2_h)$. No dependence of the Higgs contributions on $\sin\theta$
appears in the results cited in ref. ${\cite {cout}}$ ; however we agree
with the  expressions given by the same authors
in ref.  ${\cite{hewett}}$ for the limiting cases discussed in
that paper.    }~.
In the standard model only one
neutral  Higgs survives the spontaneous \mbox{electroweak}
symmetry breaking
having mass $m_{Higgs}$ whose contributions to  $\Delta k_{\gamma}$,
$\Delta Q_{\gamma}$ are expressed in terms of the ratio
$\delta={({m_{Higgs}}/M_W)}^2$; for comparison these are also given
in the {\mbox{Table I}} along with the Higgs contributions in the
supersymmetric limit.

The neutralino and chargino sectors have the most complicated structure
and will be discussed in more detail. Their weak and electromagnetic currents
are given by,
\begin{eqnarray}
{J^\mu_+}&=&{ \sum_{\alpha,i}}{\bar {\tilde Z}}_\alpha \gamma^\mu
( P_R C^R_{\alpha i}+P_L C^L_{\alpha i}) {\tilde C}_i \nonumber \\
{J_{em}^\mu}&=&{\sum_{i}} {\bar{\tilde C}}_i  \gamma^\mu   {\tilde C}_i
                                                                 \label{cur}
\end{eqnarray}
where the sums are over neutralino, $\alpha=1,2,3,4$, and/or  chargino,
$ i=1,2 $, indices and $P_{R,L}$ are the right and left handed projection
operators $(1\pm \gamma_5)/2$. ${\tilde Z}_\alpha
,\alpha=1,2,3,4$ are four component Majorana spinors eigenstates of the
symmetric neutralino mass matrix given by,
\begin{equation}
{\bf {\cal M}_N}=\left( \begin{array}{cccc}
    M_1               &     0         &g^\prime v_1 / \sqrt 2
                                                 &  -g^\prime v_2/\sqrt 2 \\
    0                 &    M_2        &  -g v_1/\sqrt 2
                                                 &   g v_2/\sqrt 2       \\
g^\prime v_1/\sqrt 2  & -g v_1/\sqrt 2&       0
                                                 &        -\mu           \\
-g^\prime v_2/\sqrt 2  & g v_2/\sqrt 2 &      -\mu
                                                 &        0
\end{array}\right)
\end{equation}
while the charginos  ${\tilde C}_i , \, i=1,2$ are Dirac fermions, mixtures
of winos $\tilde {W_+}$ and Higgsinos $\tilde {H_+}$ and they are eigenstates
of the chargino mass matrix,
\begin{equation}
{\bf {\cal M}_C}=\left(  \begin{array}{cc}
M_2      &     -g\,v_2   \\
-g\,v_1   &      \mu
\end{array}  \right)
\end{equation}

The right and left handed couplings $ C^R_{\alpha i} $ appearing in
(\ref{cur}) are given by
\begin{eqnarray}
C^R_{\alpha i}=-{\frac {1}{\sqrt 2}} O_{3 \alpha} U^{*}_{i2}-
                                     O_{2 \alpha} U^{*}_{i1} \nonumber\\
C^L_{\alpha i}=+{\frac {1}{\sqrt 2}} O_{4 \alpha} V^{*}_{i2}-
                                     O_{2 \alpha} V^{*}_{i1}
\end{eqnarray}
The real orthogonal matrix ${\bf O}$ diagonalizes ${\bf {\cal M}_N}$,
and the unitary matrices ${\bf U},{\bf V}$ diagonalize ${\bf {\cal M}_C}$
, i.e  ${\bf{O^T}}{\bf {\cal M}_N}{\bf O}=diagonal$   and
                          $ {\bf U}{\bf {\cal
M}_C}{\bf{V^{\dagger}}}=diagonal$.

The neutralino and chargino contributions to the dipole and quadrupole
moments are displayed in the Table I. For   $\Delta Q_{\gamma}$ we are in
complete agreement with the findings of ref. ${\cite {ali}}$  but for
$\Delta k_{\gamma}$  we disagree in the first term of the equation
for $\Delta k_{\gamma}$  appearing in that table. However our result
in the limit of vanishing right handed coupling,
\mbox{ $\, C^R_{\alpha i}=0\,$,}
receives a form which up to group factors is exactly the same as that of
a massive fermion familly as it should. This concists a check of its
correctness. The  $sign(m_i m_\alpha)$ in the second term of
$\Delta k_{\gamma}$ takes care of the fact that we have not committed
ourselves to a particular sign convention for the chargino and neutralino
mass eigenvalues $m_i$ and $m_\alpha$. The prefactors $F_{\alpha i}$ and
$G_{\alpha i}$ appearing in the expressions for
$\Delta k_{\gamma}$ and  $\Delta Q_{\gamma}$    are defined as
\begin{equation}
 F_{\alpha i}={|C^R_{\alpha i}|}^2+{|C^L_{\alpha i}|}^2\qquad ,\qquad
 G_{\alpha i}=( C^L_{\alpha i} \,{ {C^R_{\alpha i}}^*} +(h.c)) \nonumber
\end{equation}

In the supersymmetric limit two of the neutralinos states,  namely the photino
${\tilde \gamma}=(cos\theta_W) \,{\tilde B}+(sin \theta_W) \,{\tilde {W_3}}$
and the axino
  ${\tilde \alpha}={\frac {1}{\sqrt 2}}({\tilde \chi}_1+{\tilde \chi}_2)$
, have vanishing mass
  \footnote{${\tilde \chi}_{1,2}$ are phase rotated  Higgsino fields
   ${\tilde \chi}_{1,2}= i{\tilde H}_{1,2}$ .}.
The first combines with the photon and the second with the scalars $h_0,A$
to form supersymmetric multiplets of zero mass.
The other two linear combinations
${\tilde \alpha}^T ={\frac {1}{\sqrt 2}}(-{\tilde \chi}_1+{\tilde \chi}_2)$,
orthogonal to  ${\tilde \alpha}$, and the ``Zino"
$\tilde Z = -(sin \theta_W) {\tilde B} + (cos\theta_W) {\tilde {W_3}} $,
which is orthogonal to the photino state, have both mass $M_Z$. These two
form a new ``Zino", $\tilde \zeta$ ,  which is a Dirac fermion
describing thus four physical states of mass $M_Z$.\, This
combines with the heavy neutral $H_0$ and the Z - boson to forming a
multiplet having mass $M_Z$.
\,Also in that limit
we have two charginos, Dirac fermions of mass $M_W$, which along
with the charged Higgses $H_{\pm}$ and the W - bosons belong to
supermultiplets of mass $M_W$.
This is how supersymmetry  manifests itself in the SUSY limit
discussed previously. In that limit the contributions of fermions and
bosons of each
multiplet to $\Delta Q_{\gamma}$ cancel each other as can be seen from Table I
. Such a cancellation however does not hold for the dipole moment as is
well known $^{\cite {proy}}$.

We pass now to discuss our physical results. As we have already mentioned
in the beginning $\tan{\beta(M_Z)}$ and $m_t(M_Z)$ are considered as free
parameters; given these the top Yukawa coupling at $M_Z$ is
known and so is its value at the Unification scale
$M_{GUT} \simeq 10^{ {16}}$ GeV.
At all scales we have imposed the perturbative requirement
$h^2_2/(4\pi) \le {\cal O}(1)$. Therefore given  $m_t(M_Z)$ ,
$\tan {\beta}$ is forced to a minimum value due to the presence of an
infrared fixed point of the Yukawa coupling $^{\cite{hill,Zim}}$.
In our analysis we have limited ourselves to vanishing values
for the bottom and tau Yukawa couplings. This excludes large values of
$\tan {\beta} (\ge 15)$ ; accepting nonvanishing bottom and Yukawa couplings
produces only minor changes to the one loop dipole and quadrupole moments.
Then starting with $A_0,m_0,M_{1/2}$ at   $M_{GUT} $  we run the RGE's
of all parameters involved except those of  $B$ and  $\mu$ whose values
$B(M_Z),\mu(M_Z) $ are determined through the minimization
\mbox{conditions} (\ref{min1}),(\ref{min2}). Their values at any other scale
are found by runninng their corresponding RGE's using their near decoupling
  from the rest of the renormalization group equations. For their
determination we properly take into account the radiative corrections
to the effective potential as we have already discussed.

For a given pair of $\tan{\beta(M_Z)},m_t(M_Z)$ the parameter space
$A_0,m_0,M_{1/2}$ can be divided into three main regions,

\noindent
${\bf i)}\quad {\underline {A_0,m_0,M_{1/2} \quad comparable }} $: \\
This includes also the dilaton dominated SUSY breaking mechanism with
vanishing cosmological constant $^{\cite {kaplu}}$.  \\
${\bf ii)}\quad {\underline {M_{1/2}\gg  A_0,m_0}}$ :\\
In this case the gaugino mass is the dominant source of SUSY breaking.
This includes also the no-scale models (see second ref. in $\cite {nilles}$)
which in the physically interesting cases favor values
$ A_0=m_0=0 $ .\\
${\bf iii)}\quad {\underline { A_0,m_0 \gg M_{1/2} }}$ :\\
This includes the light gluino case.

We have explored the entire parameter space for values of the top mass
ranging from  $130$ to $180\,GeV$ and in Table II we present sample results
covering the previously discussed cases. In every case appearing in that
table the  dominant
SUSY breaking terms have been taken equal to $300\,GeV$\,. One observes
that the MSSM values for the quantities of interest, and especially for
$\Delta k_{\gamma}$, in some particular cases can be  largely different from
those of the SM but of the same order of magnitude
indicating that possible supersymmetric structure is hard to
be detected at LEP2. This behaviour  characterizes  the most of the
parameter space provided that the typical SUSY breaking scale is in the range
${\cal O}(100GeV-1 TeV)$. The  distinction between SM and MSSM predictions
can be therefore made detectable once the accuracy reaches the level of
${\cal O}(10^{-2,-3})$.

In general the contributions  of each sector separately to dipole and
quadrupole moments,
in units of  ${ {g^2} /{16{\pi^2}}}$, are as follows:

The matter fermions contributions to $\Delta k_{\gamma}$ are of  order
${\cal O}(1)$ and negative while those to  $\Delta Q_{\gamma}$ are of the
same order of magnitude but positive.

The gauge bosons contributions to $\Delta k_{\gamma}$ is of the same order
as that of the fermions but opposite in sign.\, However those to the
quadrupole moment  $\Delta Q_{\gamma}$ are almost an order of
magnitude smaller.

The supersymmetric Higgses  yield $\Delta k_{\gamma}\sim {\cal O}(1)$\,, \,
$\Delta Q_{\gamma}\sim {\cal O}(10^{-2}) $ which little deviate from the
SM predictions provided the standard model Higgs boson has a mass in the
range $\simeq 50 - 100 GeV$. Actually in the MSSM with radiative
electroweak breaking one of the Higgses, namely $h_0$ , turns out to
have a mass in the aforementioned range; this is predominantly the standard
model Higgs. The rest have large masses, of the order of the
supersymmetry breaking scale, giving therefore negligible contributions.

The squark and slepton sector is the one yielding the smallest contributions.
 These are of order  $\simeq {\cal O}(10^{-2})$ or less having therefore
negligible effect on both $\Delta k_{\gamma}$ and $\Delta Q_{\gamma}$.

It remains to discuss the effects of the charginos and neutralinos to the
dipole and quadrupole moments. In general this sector gives
$\simeq {\cal O}(10^{-2})$ to both $\Delta k_{\gamma}$ and
$\Delta Q_{\gamma}$. However in some cases and for
positive values of the parameter $\mu$, $\Delta k_{\gamma}$ can be
substantially larger, $\simeq {\cal O}(1)$. This occurs only when
$M_{1/2} \ll m_0,A_0$, case (iii),
and the solution for $\mu$ as given from the Eq.
(\ref{min1}) and (\ref{min2}) happens to allow for light chargino and
neutralino states with masses $m_{\tilde C},m_{\tilde Z}$ such that
$m_{\tilde C}+m_{\tilde Z} \approx M_W$. In such cases the values of
$\Delta k_{\gamma}$ , $\Delta Q_{\gamma}$ are enhanced and a structure is
observed \footnote{In those cases the integrations over the Feynamn
          parameter are of the form
          $\int_{0}^{1} f(t)dt/[(t-\alpha)^2+\epsilon^2]$
          with $0<\alpha<1$ and $\epsilon$ small.} (see Fig 3).
 However even for such relatively large contributions
of this sector we can not have values approaching the sensitivity
limits of LEP2.
 Only in a very limited region of the parameter space and when accidentally
$m_{\tilde C}+m_{\tilde Z} $ turns out to be almost equal to
W - boson mass,
the chargino and neutralino contributions can be very large
saturating the sensitivity limits of LEP2. We
disregard such large contributions
since they are probably outside the validity of the perturbation expansion.
 Even if it were not for that reason these cases are unnatural since they
occupy a very small portion of the available parameter space which is further
reduced if the lower experimental bounds imposed on the chargino mass
are strictly observed.

 The results presented in Table II are representative of the cases discussed
previously. To simplify the discussion
the dominant SUSY breaking parameters have been assumed equal. Notice the
large chargino/neutralino contribution to  $\Delta k_{\gamma}$ and
$\Delta Q_{\gamma}$ in the case where $M_{1/2}=80 GeV$, and
$m_0,A_0$ = $300 GeV$. Smaller values for $M_{1/2}$
can lead to even larger contributions resulting to moments
$\Delta k_{\gamma}$, $\Delta Q_{\gamma}$ which may approach the
sensitivity limits of LEP2.
However the lower experimental bound put on the chargino
mass ($\geq 45 GeV$) constraints the situation a great deal not allowing
for arbitrarilly low $M_{1/2}$ values.
In the same table the supersymmetric values $\Delta k_{\gamma}^{SUSY}$,
 $\Delta Q_{\gamma}^{SUSY}$ are also shown. For the later we know that
$\Delta Q_{\gamma}^{SUSY}=0$. Its vanishing merely serves as a check of the
correctness of the calculations.  $\Delta k_{\gamma}^{SUSY}$ is nonvanishing
however receiving the value $1.273 (g^2/{16{\pi^2}})$ for $m_t=160 GeV$.
For a comparison of our results we also display the standard
model predictions for Higgs masses ${m_{Higgs}}=50, 100$ and $ 300 GeV$.

In order to examine the behaviour of the dipole and quadrupole moments with
varying the supersymmetry breaking scale we plot in Figure 3
\, $\Delta k_{\gamma}$, $\Delta Q_{\gamma}$ as functions of  $m_0,A_0$
for the physically interesting case $M_{1/2} \ll m_0,A_0$.
To compare with the previously discussed cases we have taken
$M_{1/2}=80 GeV$ and $m_0=A_0$ ranging from $200 GeV$ to $1\, TeV$ for
both $\mu > 0$ and $\mu < 0$. Also in order to see how
sensitive are our results to the top quark mass we plot
$\Delta k_{\gamma}$, $\Delta Q_{\gamma}$  for $m_t=140 GeV$ and $m_t=160 GeV$.
The two  cases yield almost identical results showing in a clear manner
that are insensitive to the choice of the top quark mass.
They differ appreciably when we are close to values of $m_0,A_0$ for
which a dip in  $\Delta k_{\gamma}$, $\Delta Q_{\gamma}$ is
developed, occuring when  $m_t=160 GeV$ and $\mu > 0$,
due to the large neutralino and chargino contributions discussed previously.
For comparison in the same figure we have drawn
the standard model predictions for a Higgs mass equal to $100 GeV$
and $m_t=160 GeV$.
The dependence of the quantities of interest on the value of the parameter
$\tan \beta$ is rather smooth. In Figure 4 we display
$\Delta k_{\gamma}$, $\Delta Q_{\gamma}$ as functions of $\tan \beta$ for
values ranging from $2 $ to $10$. No strong dependence on $\tan \beta$ is
observed either although for $\mu > 0$ the dipole moment is appreciably
larger for small values of  $\tan \beta$.
The cases shown correspond to $m_0=A_0=300 GeV$,
$M_{1/2}=80 GeV$ and $m_t=160 GeV$. However these are representative of
the more general situation.

 Concerning our numerical results a last comment is in order. Scanning the
 parameter space we have intentionally ignored cases giving large
 contributions due to the presence of Landau singularities of our one loop
expressions. These are encountered when the masses involved take values for
which the integrands $f(t)$ develop double poles within the integration
region [0,1] of the Feynman parameter $t$.
We are aware of the fact
that near a Landau singularity the dipole and quadrupole moments can
become sizeable but these cases should be considered with a \mbox{``grain} of
salt" not corresponding to an actual physical situation. Our ignorance how
to treat these singularities forces us to keep a rather conservative view
leaving aside values of the physical masses for which we are in the vicinity
of a Landau singularity. These result to large and untrustworthy values for
the dipole and quadrupole moments which cannot be handled perturbatively.

The supersymmetric dipole and quadrupole moments though different,
in general~,
from the corresponding  standard model quantities are of the same order
of magnitude in almost the entire parameter space $m_0,A_0$  and $M_{1/2}$.
The MSSM values of these quantities are not sensitive to either top mass or
$\tan \beta$ as long as the former lies in the physical region
$140 GeV<m_t<180 GeV$. For $M_{1/2}$  small, as compared to $m_0,A_0$,
the chargino and neutralino sector may give substantial contributions
resulting to $\Delta k_{\gamma}$, $\Delta Q_{\gamma}$  approaching the
sensitivity limits of LEP2. However these cases should be
considered with some caution for they may not be allowed within the
perturbative regime; in addition these cases occupy a tiny region
in the parameter space, being therefore unnatural, and hence to be
disregarded on these grounds.
Our conclusion is that deviations from the standard model predictions
for the dipole and quadrupole moments due to supersymmetry are hard to be
observed at LEP2.
If such deviations are observed most likely these are not due to an
underlying supersymmetric structure.
\par
\vspace*{2in}
\noindent
{\bf Acknowledgements }\\  \\
The work of A.B.L. was partly supported by the EEC Science Program
SCI-CT92-0792. The work of V.C.S. is supported by the Human Capital
and Mobility programme CHRX-CT93-0319.
\newpage
%
%

\newpage
\noindent
{\bf Table Captions}

\vspace{1.cm}
{\bf Table I}:\quad Radiative corrections to the dipole $\Delta k_{\gamma}$
and quadrupole moments  $\Delta Q_{\gamma}$ in units of $g^2/(16{\pi}^2)$.
The symbols are explained in the main text.

\vspace{1.cm}
{\bf Table II}:\quad The dipole and quadrupole moments, in units of
$g^2/(16{\pi}^2)$, for the input values
\quad$m_t(M_Z),$  ${\tan \beta(M_Z),}$ $A_0,m_0$ and $M_{1/2}$ shown in
the table. For comparison we display
the standard model predictions for Higgs masses  $50$, $100$ and $300 GeV$;
the corresponding moments in the supersymmetric limit are also shown.

\vspace{2.5cm}
\noindent
{\bf Figure Captions}

\vspace{1.cm}
{\bf Figure 1}:\quad The\, $W_-W_+\gamma$\, vertex.\, Lorentz indices and
momentum assignments are as shown in the figure.

\vspace{1.cm}
{\bf Figure 2}:\quad Triangle graph (a) and its crossed (b) contributing to
the dipole and quadrupole moments. $Q,T_+,T_-$ denote electric  charge
and isospin raising and lowering operators respectively.

\vspace{1.cm}
{\bf Figure 3}:\,{\bf (a)}\quad Dipole (solid line) and Quadrupole
(dashed line) moments, in units of $g^2/(16{\pi}^2)$,
as functions of $m_0$, $A_0$ for $M_{1/2}=80 GeV$ . The cases shown are for
$m_t=140 GeV$ ($\alpha$) and $m_t=160 GeV$ ($\beta$).
 For convenience we have taken $m_0=A_0$.
The sign of the parameter $\mu$ is positive. The horizontal
lines are the standard model predictions for $m_{Higgs}$ $=100 GeV$ and
$m_t=160 GeV$.

{\bf (b)}\quad Same as in (a) for negative sign of the parameter $\mu$.

\vspace{1.cm}
{\bf Figure 4}:\quad Dipole (solid line) and Quadrupole (dashed line)
moments, in units of $g^2/(16{\pi}^2)$,
as functions of ${\tan \beta}$ for $\mu > 0\, (a)$ and
$\mu < 0\, (b)$. In both cases $A_0=m_0$ $=300 GeV$, $M_{1/2}=80 GeV$ and
$m_t=160 GeV$ .

%
%
%
%
%
%
\newpage
\begin{eqnarray*}
&&\makebox[5.5in]{\hrulefill} \\
&&\makebox[5.5in]{\bf TABLE I}\\
&&\makebox[5.5in]{\bf {Corrections\, to\, the\, Dipole\,($\Delta k_{\gamma}$)
        and\, Quadrupole\, ($\Delta Q_{\gamma}$)  moments         } }  \\
&&\makebox[5.5in]{$ {\bf
                     (\quad  in\, units\, of\,g^2/(16{\pi}^2) \quad )} $ }
\\
&&\makebox{\rule{5.5in}{.03in}} \\
&&\bf{ {Gauge} \,{ bosons}} \\
&&\makebox{\rule{5.5in}{.03in}} \\
&&    \gamma\,:\quad   \Delta k_{\gamma}= {20 \over 3}{\sin^2}{\theta_W}
             \,\,,\,\, \Delta Q_{\gamma}={4 \over 9}{\sin^2}{\theta_W}   \\
&&   Z\,:\quad    \Delta k_{\gamma}={20 \over {3R}}-{5 \over 6}+
                 {1 \over 2}  \int_{0}^{1}dt
                 {t^4+10t^3-36t^2+32t-16  \over
                 t^2+R(1-t)  }   \\
&& \quad \,\, \quad  \Delta Q_{\gamma}=({8 \over {3R}}+{1 \over 3})
  \int_{0}^{1}dt     { t^3(1-t)  \over   t^2+R(1-t)  }   \\
&&  \quad \,\, \quad  (\,\,R\,=\,{(M_Z/M_W)}^2\,)  \\
%
&&\makebox{\rule{5.5in}{.03in}}     \\
&&{\textstyle{\bf Leptons ,\,Quarks \, :}}
        \quad {f \choose f^\prime}_L \quad {\textstyle {SU(2)\,\, doublets}} \\
&&\makebox{\rule{5.5in}{.03in}}     \\
&& \Delta k_{\gamma}=
  \frac {C_g}{2}   Q_{f^\prime}  \int_{0}^{1} dt
 {  t^4+(r_f-r_{f^\prime}-1)t^3+(2r_{f^\prime}-r_f)t^2   \over
     t^2+(r_{f^\prime}-r_f-1)t+r_f  }
   -   [f\rightleftharpoons {f^\prime}]        \\
&& \Delta Q_{\gamma} =
   \frac {2C_g}{3}  Q_{f^\prime} \int_{0}^{1} dt
  {   t^3(1-t)   \over
   t^2+(r_{f^\prime}-r_f-1)t+r_f     }
   -[f\rightleftharpoons {f^\prime}]           \\
&& (r_{f,{f^\prime}}  \equiv {(m_ { f,{f^\prime}} /M_W)}^2)  \\
       \\
&&\makebox{\rule{5.5in}{.03in}}                  \\
&&{\textstyle{\bf Sleptons ,\,Squarks \, :}}
        \quad { {\tilde f} \choose {\tilde f}^\prime}_L \quad
                                        {\textstyle {SU(2) \,\, doublets}} \\
&&\makebox{\rule{5.5in}{.03in}}    \\
&& \Delta k_{\gamma}=
        - C_g  Q_{f^\prime} \sum_{i,j=1}^{2}
      {  ( {K_{i1}^{{\tilde f}} }  {K_{j1}^{{\tilde f}^\prime} }) }^2
      \int_{0}^{1}dt
   {   t^2(t-1)(2t-1+R_{{\tilde f}^\prime_j}- R_{{\tilde f}_i}) \over
      t^2+(R_{{\tilde f}^\prime_j}- R_{{\tilde f}_i}-1)t+
      R_{{\tilde f}_i}   }
      -[f\rightleftharpoons {f^\prime}]                    \\
&& \Delta Q_{\gamma}=
     - { { 2C_g} Q_{f^\prime}  \over 3   }  \sum_{i,j=1}^{2}
    {  ( {K_{i1}^{{\tilde f}} }  {K_{j1}^{{\tilde f}^\prime} }) }^2
     \int_{0}^{1}dt
   {  t^3(1-t)  \over
      t^2+(R_{{\tilde f}^\prime_j}- R_{{\tilde f}_i}-1)t+
      R_{{\tilde f}_i}   }
     -[f\rightleftharpoons {f^\prime}]                    \\
&&   \\
&&   R_{ {\tilde f}_i , {{\tilde  f}^\prime}_i }
     \equiv {(m_ { {\tilde f}_i , {{\tilde  f}^\prime}_i }  /M_W)}^2
 ;\quad m_ { {\tilde f}_i , {{\tilde  f}^\prime}_i } \quad are\,
 \,sfermion\,\,masses. \quad    \\
&& In\,\,the\,\,SUSY\,\,limit\,\,{\bf K}^ { {\tilde f},{\tilde f}^\prime } \,
\, become\,\,
unit\,\,matrices
\,\,and\,\, R_{{\tilde f}_{1,2}}=r_f, \,\,
  R_{  {{\tilde f}^\prime}_{1,2}  }=r_{f^\prime}  \\
&&\makebox{\rule{5.5in}{.03in}} \\
&&\makebox [5.5in][r]{\bf (continued)} \\
\\
\\
&&{\bf{  \underline{ TABLE \quad I}  \quad (continued)  }   } \\
&&\makebox{\rule{5.5in}{.03in}}      \\
&&{\bf {Higgses\,: \,(H_{\pm},\,H_0,\,h_0,\,A)} }\\
&&\makebox{\rule{5.5in}{.03in}}     \\
&& \\
&& a)\underline { broken\,\,\,SUSY}   \\
&& \\
&&A\;:\qquad \Delta k_{\gamma}=D_2(R_A,R_+) \quad , \quad
            \Delta Q_{\gamma}=Q(R_A,R_+) \\
&&h_0\,:\qquad \Delta k_{\gamma}\:=\: \sin^2 \theta \: D_1\,(R_h)\:+\:
                                \cos^2 \theta \: D_2(R_h,R_+)  \\
&&\quad \qquad \hspace*{4.5mm} \Delta Q_{\gamma}\:=\: \sin^2 \theta \:
                             Q(R_h,1)\:+\: \cos^2 \theta \:Q(R_h,R_+)   \\
&&H_0\,:\quad \, \; \; As \,\, in\quad h_0\quad with \quad R_h \rightarrow R_H
\quad and
       \quad  \sin^2 \theta \rightleftharpoons  \cos^2 \theta   \\
&& \\
&&    R_a\equiv{(m_a/M_W)}^2 \quad a=h_0,H_0,A,H_{\pm} \\
&& \\
&& b)\underline { exact\,\,\,SUSY}   \\
&& \\
&&A\,,h_0\,:\qquad  \Delta k_{\gamma}=- {1\over 6} \, ,\,+{11\over 6}
   \quad ,\quad        \Delta Q_{\gamma}={1\over 18}\, , \, {1\over 18}    \\
&&H_0 \, \quad  : \qquad  \Delta k_{\gamma}={1\over 6}+
           {1\over 2} \int_{0}^{1} dt  {t^4-2t^3 \over t^2+R(1-t) }
 \quad ,\quad     \Delta Q_{\gamma}=
           {1\over 3} \int_{0}^{1} dt {t^3(1-t)  \over t^2+R(1-t)  }  \\
&&\quad\quad  (\,\,R\,=\,{(M_Z/M_W)}^2\,)  \\
&&   \\
&&c)\underline {Standard\,\,Model\,\,Higgs\,\,contribution}\\
&& \\
&&\quad\quad   \Delta k_{\gamma}=D_1\,(\delta) \quad , \quad
               \Delta Q_{\gamma}=Q(\delta,1) \qquad
   (\,\,\delta\,=\,{(m_{Higgs}/M_W)}^2\,)            \\
&& \\ \\ \\
&& \rhd  \quad \quad  D_1(r)\equiv{1\over 2} \int_{0}^{1} dt
        { 2t^4+(-2-r)t^3+(4+r)\,t^2   \over
        t^2+r(1-t)   }      \\
&&\qquad \quad   D_2(r,R)\equiv{1\over 2} \int_{0}^{1} dt
        { 2t^4+(-3-r+R)t^3+(1+r-R)t^2   \over
        t^2+(-1-r+R)t+r   }  \\
&&\qquad \quad   Q(r,R)\equiv{1\over 3} \int_{0}^{1} dt
        { t^3(1-t) \over
         t^2+(-1-r+R)t+r }       \\
&&  \\
&&\makebox{\rule{5.5in}{.03in}} \\
&&\makebox [5.5in][r]{\bf (continued)} \\  \\  \\
&&{\bf{  \underline{ TABLE \quad I}  \quad (continued)  }   } \\
&&\makebox{\rule{5.5in}{.03in}} \\
&&{\bf  {Neutralinos \, (\,{\tilde Z}_\alpha ,\, \alpha=1...4)  }}
 \quad {\bf  { Charginos \,  (\, {\tilde C}_i,\,i=1,2)   }}  \\
&&\makebox{\rule{5.5in}{.03in}} \\
&&   \\
&& a)\underline { broken\,\,\,SUSY}   \\
&& \Delta k_{\gamma}=- \sum_{i,\alpha} F_{\alpha i} \int_{0}^{1} dt
      {t^4+(R_{\alpha}-R_i-1)t^3+(2R_i-R_{\alpha})t^2  \over
      t^2+(R_i-R_{\alpha}-1)t+R_{\alpha}  }   \\
&&\qquad \quad +\sum_{i,\alpha}\,sign(m_im_\alpha)\,\,G_{\alpha i}\,
   \sqrt{R_{\alpha}R_i} \int_{0}^{1} dt {4t^2-2t \over
      t^2+(R_i-R_{\alpha}-1)t+R_{\alpha}  } \\
&& \Delta Q_{\gamma}=-{4 \over 3}\sum_{i,\alpha} F_{\alpha i} \int_{0}^{1} dt
        { t^3(1-t) \over t^2+(R_i-R_{\alpha}-1)t+R_{\alpha}  }  ,   \\
&& \quad \quad       (\, R_{\alpha ,i}\equiv{(m_{\alpha,i}/M_W)}^2\, ) \\   \\
&&   \\
&& b)\underline { exact\,\,\,SUSY}   \\
&&   \\
&&   {\tilde \gamma}\,:\quad \Delta k_{\gamma}=- {8 \over 3}{\sin^2}{\theta_W}
   \quad,\quad  \Delta Q_{\gamma}=-{4 \over 9}{\sin^2}{\theta_W}   \\
&& {\tilde \alpha}\,:\quad  \Delta k_{\gamma}=- {2 \over 3}
  \quad\quad \quad\quad,\quad  \Delta Q_{\gamma}=-{1 \over 9}  \\
&& {\tilde \zeta}\,:\quad  \Delta k_{\gamma}={1 \over 6}-{8 \over 3R}
       -  \int_{0}^{1} {t^4+3t^3-15t^2+12t-4 \over
                        t^2+R(1-t) } \\
&& \quad  \quad\,\,  \Delta Q_{\gamma}=-({2 \over 3}+ {8 \over 3R})
       \int_{0}^{1}  {t^3(1-t) \over   t^2+R(1-t) } \\
&&  \quad\,\quad (R\,=\,{(M_Z/M_W)}^2\,)  \\
&&    \\
&&\makebox{\rule{5.5in}{.01in}}
\end{eqnarray*}
%
%
\newpage
\vspace*{3cm}
\oddsidemargin=-4mm
\begin{center}
\begin{tabular}{|c||c|c||c|c|} \hline
\multicolumn{5}{|c|}{  \bf {TABLE II}  }  \\  \hline
\multicolumn{5}{|c|}{$ m_t=160,\tan{\beta}=2$ } \\
\multicolumn{5}{|c|}{$ A_0,m_0,M_{1/2}\quad:\quad300, 300, 300 \quad
 [0, 0, 300]\quad(300, 300, 80) $}\\ \hline
       &\multicolumn{2}{c||}{$\Delta k_{\gamma}$ }
       & \multicolumn{2}{c|}{$\Delta Q_{\gamma}$ } \\ \hline \hline
   & $\mu > 0$  & $\mu < 0$ & $\mu > 0$  & $\mu < 0$  \\ \hline
$q,l$ & \multicolumn{2}{c||}{-1.973}
           & \multicolumn{2}{c|}{1.922}        \\ \hline
$W,\gamma,Z$ & \multicolumn{2}{c||}{1.179}
               & \multicolumn{2}{c|}{0.235}     \\ \hline
$h_0,H_{\pm,0},A$ & \multicolumn{2}{c||}{.945[.945](.946)}
          & \multicolumn{2}{c|}{.029[.028](.028)}   \\ \hline
${\tilde q},{\tilde l}$&-.004[-.024](.009)&-.013[-.032](-.035)
                      &.009[.019](.027)&.009[.019](.025)  \\ \hline
${\tilde Z},{\tilde C}$&.013[.015](.697)&-.003[-.005](.026)
                     &-.016[-.017](-.592)&-.014[-.014](-.170)  \\ \hline
        &  &  &  &  \\
Total &.159[.143](.859) & .135[.115](.143)
      &2.178[2.188](1.621)  &2.180[2.190](2.041)  \\
         &  &  &  &  \\  \hline
        &\multicolumn{2}{c||}{ }   &\multicolumn{2}{c|}{ }  \\
SUSY  &\multicolumn{2}{c||}{   ${\Delta k_{\gamma}}^{SUSY}\quad=\quad1.237$  }
    &\multicolumn{2}{c|} {   ${\Delta Q_{\gamma}}^{SUSY}\quad=\quad 0.   $ }
              \\
limit & \multicolumn{2}{c||}{ }   &\multicolumn{2}{c|}{ }  \\  \hline
        &\multicolumn{4}{c|}{ }   \\
Standard  &\multicolumn{4}{c|}{ ${\Delta k_{\gamma}}^{SM}
                \quad=\quad   .188,\,-.106,\,-.449 $ }  \\
Model  &\multicolumn{4}{c|}{ ${\Delta Q_{\gamma}}^{SM}
                \quad=\quad   2.186,\,2.174,\,2.161 $ }  \\
            &\multicolumn{4}{c|}{ }   \\   \hline
\end{tabular}
\end{center}
\oddsidemargin=0mm
\newpage
%
\begin{center}
\begin{picture}(25000,20000)
\THICKLINES
\bigphotons
\drawline\photon[\S\REG](12000,18000)[7]
\global\advance\pfronty by 500
\put(\pfrontx,\pfronty){${\gamma_\mu}$}
\put(13000,15000){\vector(0,-1){3000}}
\put(13000,16000){$2Q$}
\put(12000,9100){\circle{5000}}
\drawline\photon[\NE\REG](6000,3500)[7]
\global\advance\pfrontx by -1000
\global\advance\pfronty by -1000
\put(\pfrontx,\pfronty){${W_{\alpha}^+}$}
\put(5000,5000){\vector(1,1){3000}}
\put(3000,7000){$\Delta-Q$}
\drawline\photon[\NW\REG](18000,3500)[7]
\global\advance\pfrontx by 500
\global\advance\pfronty by -1000
\put(\pfrontx,\pfronty){${W_{\beta}^-}$}
\put(18000,5000){\vector(-1,1){3000}}
\put(18000,7000){$-\Delta-Q$}
\end{picture}
\end{center}
\vspace*{.3cm}
\par
{\hspace*{6.3cm}}
{\bf {Figure 1}}
\vspace{1.3cm}
%
\begin{center}
\begin{picture}(28000,30000)
\THICKLINES
\bigphotons
\drawline\photon[\S\REG](4500,25000)[7]
\global\advance\pfronty by 1500
\put(\pfrontx,\pfronty){$\gamma(2Q)$}
\global\advance\photonbackx by 500
\put(\photonbackx,\photonbacky){$\bf Q$}
%
\drawline\fermion[\SW\REG](\pbackx,\pbacky)[6000]
\drawarrow[\NE\ATBASE](\pmidx,\pmidy)
\drawline\photon[\S\REG](\fermionbackx,\fermionbacky)[6]
\global\advance\fermionbackx by -2000
\put(\fermionbackx,\fermionbacky){$\bf {T_+} $}
\global\advance\fermionbackx by 2000
\global\advance\photonbacky by -2500
\global\advance\photonbackx by -1000
\put(\photonbackx,\photonbacky){$ {W_{\alpha}^+}(p) $}
\drawline\fermion[\E\REG](\fermionbackx,\fermionbacky)[8485]
\drawarrow[\W\ATBASE](\pmidx,\pmidy)
\drawline\fermion[\NW\REG](\pbackx,\pbacky)[6000]
\drawarrow[\SE\ATBASE](\pmidx,\pmidy)
\drawline\photon[\S\REG](\fermionfrontx,\fermionfronty)[6]
\global\advance\fermionfrontx by 1000
\put(\fermionfrontx,\fermionfronty){$\bf {T_-} $}
\global\advance\fermionbackx by -1000
\global\advance\photonbacky by -2500
\global\advance\photonbackx by -1000
\put(\photonbackx,\photonbacky){$ {W_{\beta}^-}(p^\prime) $}
\drawline\photon[\S\REG](22000,25000)[7]
\global\advance\pfronty by 1500
\put(\pfrontx,\pfronty){$\gamma(2Q)$}
\global\advance\photonbackx by 500
\put(\photonbackx,\photonbacky){$\bf Q$}
\drawline\fermion[\SW\REG](\pbackx,\pbacky)[6000]
\drawarrow[\NE\ATBASE](\pmidx,\pmidy)
\drawline\photon[\SE\REG](\fermionbackx,\fermionbacky)[11]

\global\advance\fermionbackx by -2000
\put(\fermionbackx,\fermionbacky){$\bf {T_-} $}
\global\advance\fermionbackx by 2000

\global\advance\photonbackx by -1500
\global\advance\photonbacky by -2000
\put(\photonbackx,\photonbacky){$ {W_{\beta}^-}(p^\prime) $}

\drawline\fermion[\E\REG](\fermionbackx,\fermionbacky)[8485]
\drawarrow[\W\ATBASE](\pmidx,\pmidy)
\drawline\fermion[\NW\REG](\pbackx,\pbacky)[6000]
\drawarrow[\SE\ATBASE](\pmidx,\pmidy)
\drawline\photon[\SW\REG](\fermionfrontx,\fermionfronty)[6]
\global\advance\photonbackx by -800
\global\advance\photonbacky by -800
\drawline\photon[\SW\REG](\photonbackx,\photonbacky)[4]
\global\advance\fermionfrontx by 1000
\put(\fermionfrontx,\fermionfronty){$\bf {T_+} $}
\global\advance\fermionbackx by -1000

\global\advance\photonbackx by -2000
\global\advance\photonbacky by -2000
\put(\photonbackx,\photonbacky){$ {W_{\alpha}^+}(p) $}
\put(4500,1500){(a)}
\put(21500,1500){(b)}
\end{picture}
\end{center}
\vspace*{.3cm}
\par
{\hspace*{6.3cm}}
{\bf {Figure 2}}
\newpage
\vspace*{8cm}
\begin{center}
\bf {Figure 3}
\end{center}
\vspace{10cm}
\begin{center}
\bf {Figure 4}
\end{center}
\end{document}